\documentclass[11pt]{article}
\usepackage{verbatim}       
\usepackage{amsfonts}       
\usepackage{amsmath}        
\usepackage{bm}         
\usepackage{cite}       

\def\be{\begin{equation}}
\def\ee{\end{equation}}
\def\bea{\begin{eqnarray}}
\def\eea{\end{eqnarray}}
\def\k{\kappa}
\def\a{\alpha}
\def\d{\delta}
\def\q{\theta}
\def\w{\omega}
\def\ep{{\epsilon}}
\def\R{{\mathbb R}}
\def\pa{\partial}
\def\wn{w}
\def\Vol{{\cal V}}
\def\e{{\varepsilon}}

\begin{document}
\begin{flushright}
NSF-KITP-07-59 \\
arXiv:0704.1160
\end{flushright}

\begin{center}
\vspace{1cm} { \LARGE {\bf Hall conductivity from dyonic black holes}}

\vspace{1.1cm}

Sean A. Hartnoll and Pavel K. Kovtun

\vspace{0.8cm}

{\it KITP, University of California\\
     Santa Barbara, CA 93106-4030, USA }

\vspace{0.8cm}

{\tt hartnoll@kitp.ucsb.edu, kovtun@kitp.ucsb.edu} \\

\vspace{2cm}

\end{center}

\begin{abstract}
\noindent
A class of strongly interacting 2+1 dimensional conformal field
theories in a transverse magnetic field can be studied using the
AdS/CFT duality. We compute zero momentum hydrodynamic response
functions of maximally supersymmetric 2+1 dimensional $SU(N)$
Yang-Mills theory at the conformal fixed point, in the large $N$
limit. With background magnetic field $B$ and electric charge
density $\rho$, the Hall conductivity is found to be $\rho/B$. The
result, anticipated on kinematic grounds in field theory, is
obtained from perturbations of a four dimensional AdS black hole
with both electric and magnetic charges.

\end{abstract}

\pagebreak
\setcounter{page}{1}

\section{Introduction}

The Hall effect is a fundamental property of a conducting medium
subject to an external magnetic field. Upon application of an
electric field, current flows in a direction orthogonal to both
the electric and magnetic fields. Therefore the conductivity
tensor, defined by $j_a = \sigma_{ab} E_b$, acquires off diagonal
entries.

In a Lorentz invariant theory, the d.c. conductivity in the
presence of an external magnetic field is completely determined by
boost invariance. Suppose there is a magnetic field ${\bm B}$ in
the lab frame. Consider a frame moving with small velocity $- {\bm
v}$ with respect to the lab frame. In this frame there is a
current ${\bm j} = \rho {\bm v}$, where $\rho$ is the charge
density of the medium in the lab frame, and an electric field
\be\label{eq:lorentz}
{\bm E} = - {\bm v} \times {\bm B} = -
\frac{1}{\rho} {\bm j} \times {\bm B} \,.
\ee
If the magnetic field is ${\bm B} = (0,0,B)$, then from
(\ref{eq:lorentz}) we have that in the $xy$ plane
\be\label{eq:classical}
\sigma_{xy} = - \sigma_{yx} = \frac{\rho}{B}
\qquad \textrm{and} \qquad \sigma_{xx} = \sigma_{yy} = 0 \,.
\ee
Thus the conductivity tensor is antisymmetric:
its off diagonal components are precisely the Hall conductivity.

From a microscopic point of view, $\sigma_{ab}$ can be
evaluated using the Kubo formula which
relates electrical conductivity to
the current-current retarded Greens function,
evaluated in the thermal equilibrium state
\begin{equation}
   \sigma_{ab} = - \lim_{\w \to 0} \frac{\textrm{Im}\,G_{ab}^R(\w)}{\w}\,.
\label{eq:Kubo-formula}
\end{equation}
The validity of the Kubo formula only relies on linear response;
in particular, it does not assume that the physical system is
composed of weakly interacting quasiparticles. Thus, for strongly
interacting systems without a quasiparticle description, equation
(\ref{eq:Kubo-formula}) provides a first-principles way to
evaluate the conductivity. We note that Kubo formulae are modified
in the presence of an external magnetic field. The expression
(\ref{eq:Kubo-formula}) remains valid however.

Motivated by understanding charge transport at
quantum critical points \cite{ssbook},
in this paper we will study conductive properties
of strongly interacting 2+1 dimensional
conformal field theories (CFTs) in the presence of an external
magnetic field.
A particular field theory to which our discussion applies
is the infrared conformal fixed
point of maximally supersymmetric $SU(N)$ Yang Mills theory at
large $N$. This CFT has eight supersymmetries and a global $SO(8)$
R symmetry group. The magnetic field we will turn on belongs to a
$U(1)$ subgroup of $SO(8)$. The theory describes the low energy
dynamics of M2 branes in M theory. However, as emphasised in the
recent work \cite{Herzog:2007ij}, it is our hope that the study of
this special theory may shed light upon general aspects of
strongly coupled CFTs in 2+1 dimensions that are of interest in
quantum critical phenomena. In fact, our results will apply to any
CFT with an AdS/CFT dual that may be truncated to Einstein-Maxwell
theory on $AdS_4$.

The M2 brane theory is tractable in the large $N$ limit because
the AdS/CFT correspondence \cite{MAGOO} provides a dual
gravitational description of the system which may be treated
classically in this limit. Furthermore, it has been understood how
the correspondence may be used to extract hydrodynamic
coefficients that describe the large scale and late time behaviour
of finite temperature field theories
\cite{Policastro:2001yc, Son:2002sd, Policastro:2002se}. In 3+1
dimensions, that work has led to a conjectured universal lower
bound on the ratio of shear viscosity to entropy density
\cite{Kovtun:2004de} and also to unanticipated connections with
the fireball created at the Relativistic Heavy Ion Collider
\cite{Steinberg:2007iw}. In contrast, the implications of the
AdS/CFT correspondence for 2+1 dimensional hydrodynamics have so
far been less developed, although see
\cite{Herzog:2002fn, Herzog:2003ke, Saremi:2006ep, Herzog:2007ij}.

The behaviour under a magnetic field is a basic probe of
interacting 2+1 dimensional systems.
In the following section we show how the M2 brane
theory may be placed in a background magnetic field by considering
a dual magnetically charged black hole in $AdS_4$. We go on to
describe the thermodynamics and some of the hydrodynamic response
functions of this theory.
The AdS/CFT duality maps fluctuations of conserved currents in
the 2+1 dimensional CFT to fluctuations of gauge fields
in the background of the 3+1 dimensional black hole,
and provides a recipe for computing current-current correlation functions.
We will use the Kubo formula (\ref{eq:Kubo-formula}) to compute the Hall
conductivity from perturbations of the black hole spacetime and
recover precisely (\ref{eq:classical}).

\section{Dyonic black hole in $AdS_4$}

In this section we describe the four dimensional spacetime dual to
the M2 brane CFT on $\R^{1,2}$. We are interested in the system at
finite temperature and with a background magnetic field. Finite
temperature is realised in AdS/CFT by allowing the spacetime to
contain a black hole \cite{Witten:1998zw}. We will shortly explain
that a background magnetic field is obtained by allowing the black
hole to carry a magnetic charge.

The fact that our CFT is relativistic implies that it has the same
number of excitations with positive and negative charges. Under
applied magnetic and electric fields, the charges will create
opposite currents that cancel and there will be no Hall
conductivity. To avoid this scenario, we need to consider the CFT
in a state with a net charge density. We will recall below that
this is obtained by requiring that the black hole carry an
electric charge.

In summary: the background we require is a dyonic black hole in
$AdS_4$, with both electric and magnetic charge. Such black holes
have been known for some time \cite{Romans:1991nq}. Precisely four
spacetime dimensions are necessary for a point source to be both
magnetically and electrically charged.

The full supergravity context is eleven dimensional supergravity
on $AdS_4 \times S^7$. This theory may be consistently truncated
to Einstein-Maxwell theory on $AdS_4$. For details, see for
instance \cite{Herzog:2007ij}. In this reduction, the Maxwell
field originates via the Kaluza-Klein mechanism as a $U(1)$
subgroup of the $SO(8)$ symmetry group of the full background. The
action for Einstein-Maxwell theory with a negative cosmological
constant $-1/L^2$ is
\be\label{eq:theaction}
I = \frac{2}{\k^2_{4}} \int d^4x \sqrt{-g}
\left[-\frac{1}{4} R + \frac{L^2}{4} F_{\mu\nu} F^{\mu\nu} -
\frac{3}{2} \frac{1}{L^2} \right] \,,
\ee
which implies equations of motion
\begin{subequations}
\label{eq:eom}
\bea
R_{\mu\nu} & = & 2 L^2 F_{\mu\sigma} F_\nu{}^\sigma -
\frac{L^2}{2} g_{\mu\nu} F_{\sigma\rho} F^{\sigma\rho} -
\frac{3}{L^2} g_{\mu \nu} \,,
\label{eq:einstein-eqs} \\
\nabla_\mu F^{\mu \nu} & = & 0 \,.
\label{eq:maxwell-eqs}
\eea
\end{subequations}
We have included the overall normalisation of the action coming
from the reduction of eleven dimensional supergravity, which in
terms of the field theory $N$ is (see e.g. \cite{Herzog:2007ij})
\be
\frac{2 L^2}{\k^2_{4}} = \frac{\sqrt{2} N^{3/2}}{6 \pi} \,.
\ee
Our background metric will be a black hole in $AdS_4$ with planar
horizon
\be
\frac{1}{L^2} ds^2 = \frac{\a^2}{z^2} \left[-f(z) dt^2 + dx^2 + dy^2\right] +
\frac{1}{z^2} \frac{dz^2}{f(z)} \,.
\label{eq:metric}
\ee
We will take the black hole to carry both electric and magnetic
charge
\be\label{eq:Ffield}
F = h \a^2 dx \wedge dy + q \a dz \wedge dt \,,
\ee
which implies that
\be
f(z) = 1 + (h^2 + q^2) z^4 - (1 + h^2 + q^2) z^3 \,.
\label{eq:f}
\ee
This solution may be obtained, for instance, by taking the planar limit
of the expressions in \cite{Romans:1991nq}.
One can explicitly check that Eqs.~(\ref{eq:metric}) -- (\ref{eq:f})
solve the equations of motion (\ref{eq:eom}).
Without loss of generality we have scaled the coordinates so that
the horizon is at $z=1$. The AdS
asymptopia is at $z \to 0$. 
The parameter $\a$ has the dimensions of mass, and determines the
temperature of the black hole through
\be\label{eq:temperature}
T = \frac{\a (3 - h^2 - q^2)}{4 \pi} \,.
\ee
Note that for a regular horizon we have $3-h^2-q^2 > 0$.
This inequality defines the allowed range
of $h$ and $q$. It will not, however, restrict the ranges of the dual
variables $T,B$ and $\rho$. The extremal zero temperature limit is
achieved by taking $h^2+q^2 \to 3$,
with $\a$ fixed. The relation of $\a$ to the mass of the
black hole is given in the following section.

The dual field theory to this spacetime is the low energy theory
living on $N$ M2 branes with worldvolume $\R^{1,2}$. To understand
the implications of the bulk Maxwell field it is convenient to
consider a potential giving $F=dA$,
\be\label{eq:Afield}
A = h \a^2 x dy + q \a z dt \,.
\ee
We see that the magnetic term remains finite at the AdS boundary,
$z \to 0$, whereas the electric term goes to zero. These different
falloffs lead to differing dual interpretations for the charges
\cite{Klebanov:1999tb}. Both falloffs of a Maxwell field in $AdS_4$
are normalisable, and we can choose which one to interpret as being dual
to a VEV \cite{Klebanov:1999tb}. We will make the standard choice in which the faster
falloff is dual to a VEV.

The magnetic term in the potential (\ref{eq:Afield}) has a slower falloff,
remaining finite at the AdS
boundary, and hence corresponds to an external magnetic field for
a gauged $U(1)$ subgroup of the $SO(8)$ R symmetry group of the
theory. Such modes are usually considered as adding
the term $A^0_\mu J^\mu$ to the field theory Lagragian, where
$A^0_\mu$ is the boundary value of the field and $J^\mu$ is the
dual $U(1)$ current. This is the same as gauging the $U(1)$
symmetry by adding the background field $A^0_\mu$, which in this
case is magnetic. The strength of the magnetic field in the field
theory is $B = h \a^2$, as we can read off from taking $z \to 0$
in the expression for the bulk field strength (\ref{eq:Ffield}).

The electric term in the potential has a faster falloff, and therefore
does not correspond to a background field in the dual theory.
Instead, it fixes the electric charge density of the state in the
field theory to be
\be\label{eq:charge1}
\rho \equiv \langle J^t \rangle = \frac{\delta I}{\delta A^{0}_t}
= - \frac{\sqrt{2} N^{3/2}}{6 \pi} q \a^2.
\ee
Here $J^t$ is the charge density operator for the same $U(1)$
subgroup of the R symmetry as before. We will rederive this
expression for the charge density from thermodynamic
considerations in the following section.

Finally, there is in fact a term missing in the potential
(\ref{eq:Afield}). In order for the potential to be regular at the
horizon of the black hole, $A_t$ must vanish there.%
\footnote{
   One way to see this is to look at the Euclidean black hole
   solution with imaginary time direction $\tau$ compactified to a
   circle. The radius of the circle shrinks to zero at the
   horizon, implying that $A_\tau$ must vanish there.
}
This requires
that we add to (\ref{eq:Afield}) the pure gauge term $-q \a dt$.
This term remains finite as $z\to 0$ and has the dual interpretation of
adding a chemical potential for the electric charge, $\mu = - q
\a$, to the field theory. This is the chemical potential
corresponding to the electric charge density $\rho$. Our
computation of the conductivity will only depend on the bulk field
strength (\ref{eq:Ffield}) and not on this pure gauge term.

\section{Thermodynamics of the grand canonical ensemble}

Before going on to compute hydrodynamic correlators, it is useful
to summarise the thermodynamic properties of the black holes we
have just described. These characterise the dual field theory in
thermal equilibrium.

We describe the thermodynamics of the grand canonical ensemble,
where the chemical potential $\mu = - q \a$ is kept fixed. We also
keep fixed the magnetic field $B = h \a^2$, this is treated as a
parameter of the system rather than a thermodynamic variable. We
will see that this is a consistent treatment.

We give the results in terms of the bulk spacetime variables
$q,h,\alpha$. There are various ways of computing thermodynamic
quantities from black holes. The most elegant is holographic
renormalisation, in which the action is regularised using a
counterterm boundary action \cite{Henningson:1998gx, Balasubramanian:1999re}.
No new counterterms are needed due to the Maxwell field, as the $F^2$
term in the action falls off sufficiently quickly near the
boundary. The renormalised action is given by substracting a
boundary term from the bulk action
\be
I_{\textrm{ren.}} = I - \frac{1}{\k^2_4} \int d^3 x \sqrt{-\gamma}\,\theta
 - \frac{2}{\k^2_{4}} \frac{1}{L} \int d^3 x \sqrt{- \gamma} \,,
\label{eq:I-ren}
\ee
where $\gamma$ is the boundary metric.
We have also included the Gibbons-Hawking boundary term.
Here $\theta=\gamma^{\mu\nu}\theta_{\mu\nu}$
is the trace of the extrinsic curvature
$\theta_{\mu \nu} = - \frac12(\nabla_{\mu} n_{\nu}+ \nabla_\nu n_\mu)$,
with $n$ an outward directed unit normal vector to the boundary.

The thermodynamic potential is given by the renormalised action
$I_{\textrm{ren.}}$ evaluated on the solution times the
temperature (\ref{eq:temperature})
\be
\Omega = T I_{\textrm{ren.}} = \frac{\sqrt{2} N^{3/2}}{6 \pi} \frac{\Vol \a^3}{4}
\left(-1 - q^2 + 3 h^2 \right) \,.
\ee
Here $\Vol = \int\! dx\, dy$ is the spatial volume.

The renormalised energy momentum tensor
of the black hole is \cite{Balasubramanian:1999re}
\be
\frac{1}{L^3} \langle T^{\mu \nu}_\text{b.h.} \rangle = \frac{2}{\sqrt{-\gamma}}
\frac{\delta I_{\textrm{ren.}}}{\delta \gamma_{\mu \nu}}
= \frac{\sqrt{2} N^{3/2}}{6 \pi} \frac{1}{2} \left[\q^{\mu \nu} -
  \theta \gamma^{\mu \nu} - \frac{2}{L} \gamma^{\mu \nu} \right]  \,.
\ee
This is related to the field theory energy momentum tensor by
$\langle T^{\mu\nu}\rangle = (\a/z)^5 \langle
T^{\mu\nu}_\text{b.h.} \rangle$. Thus we can obtain the energy
\be
E = \int d^2 x \langle T^{\mu \nu}_\text{b.h.} \rangle k_{\mu}
\xi_{\nu} \sqrt{\sigma} = \frac{\sqrt{2} N^{3/2}}{6 \pi}
\frac{\Vol \a^3}{2}
\left(1 + q^2 + h^2 \right) \,.
\ee
The integral is over the $\R^2$ at spatial infinity $z \to 0$, $k$
is a unit vector normal to the spatial hypersurface
$t=\textrm{const.}$, $\xi$ is the Killing vector $\pa_t$ and
$\sqrt{\sigma} = (\gamma_{xx} \gamma_{yy})^{1/2}$ is the volume
element of the $\R^2$. The entropy is given by the area of the
horizon times a standard normalisation to be
\be
S = \frac{\sqrt{2} N^{3/2}}{6} \Vol \a^2 \,.
\ee
The total electric charge may be computed
by varying the free energy with respect to the chemical potential
\be\label{eq:charge}
Q =
  - \left(\frac{\pa \Omega}{\pa\mu}\right)_{T,B} =
  \frac{\sqrt{2} N^{3/2}}{6 \pi} \Vol \a \mu \,.
\ee
This expression agrees with our previous result (\ref{eq:charge1}).
It will be useful later to define the energy, entropy and charge
densities
\be\label{eq:densities}
\e = \frac{E}{\Vol} \,, \qquad s = \frac{S}{\Vol} \,, \qquad \rho =
\frac{Q}{\Vol} \,.
\ee
Finally, the pressure in the grand canonical ensemble is simply
given by
\be
\Omega = - P \Vol \,.
\ee
Note that in a magnetic field, $P$ differs from $\langle T_{xx}
\rangle$ by a term proportional to the magnetisation.
A check of the formulae we have given in this section is
that they satisfy the required thermodynamic relation
\be
\Omega = E - T S - \mu Q \,.
\ee
The fact that this relation holds without needing to add a term
for the magnetic charge shows that our treatment of $h$ as a
constant external parameter is consistent.

Analogously to the five dimensional example in \cite{ss-r}, we check
the local thermodynamic stability of the system by
considering the equation of state $\e(s,\rho)$. In the grand
canonical ensemble, the condition for stability is that $\det
\left[ \pa^2_{s\rho} \e(s,\rho) \right] > 0$. From the formulae
above it follows that
\begin{equation}
\e(s,\rho) = \frac{6^{1/2}}{2^{1/4} N^{3/4}}
\frac{s^{3/2}}{2\pi} \left[1 + \frac{\rho^2 \pi^2}{s^2}
+ \frac{\tilde B^2 \pi^2}{s^2} \right] \,,
\end{equation}
where $\tilde B = B \sqrt{2} N^{3/2}/6\pi$. It is easily checked
that the condition for the determinant to be positive is that $3 +
3 h^2 + q^2 > 0$. This is certainly true and therefore the system
is locally thermodynamically stable at all temperatures, charges
and values of the background magnetic field.

\section{Fluctuations and action}
\subsection{Equations of motion}

We are aiming to compute correlators of the boundary current
operators $J_x$ and $J_y$, dual to the components of the bulk
Maxwell potential $A_x$ and $A_y$. The AdS/CFT dictionary requires
that we consider fluctuations of these fields about the black hole
background. In order to extract the conductivity, it will suffice
to work at zero momentum in the $x$ and $y$ directions. [It is
consistent to do so because only the background field strength
(\ref{eq:Ffield}) enters the equations, not the background
potential (\ref{eq:Afield}). Thus, translation invariance is
maintained.] That is, the perturbations are taken to be
independent of $x$ and $y$. In this case, it turns out that the
gauge field fluctuations source fluctuations in the metric
components $g_{tx}$ and $g_{ty}$ and no others. By linearising the
Einstein-Maxwell equations about the background we obtain the
following equations for the fluctuations: From the Maxwell
equations (\ref{eq:maxwell-eqs}) we find
\begin{subequations}
\label{eq:maxwell-eqs-2}
\bea
f (f A_x')' + \wn^2 A_x + i \wn h G_y + q f G_x' & = & 0 \,,
\\
f (f A_y')' + \wn^2 A_y - i \wn h G_x + q f G_y' & = & 0 \,.
\eea
\end{subequations}
In these equations $G_x = g_{tx} \a^{-1} z^2$, and similarly for
$G_y$. Prime denotes differentiation with respect to $z$. The time
dependence is taken to be $e^{-i \w t}$ for all fields. We have
also introduced the dimensionless frequency $\wn\equiv\w\a^{-1}$.

The Einstein equations (\ref{eq:einstein-eqs}) give
\begin{subequations}
\label{eq:einstein}
\bea
f ( G_y'/4 z^2)' - h^2 G_y + i \wn h A_x + q f A_y' = 0 \,,
\\
f ( G_x'/4 z^2)' - h^2 G_x - i \wn h A_y + q f A_x' = 0 \,,
\\
i \wn G_y'/4 z^2 + h f A_x' + i\wn q A_y + h q G_x = 0 \,,
\\
i \wn G_x'/4 z^2 - h f A_y' + i\wn q A_x - h q G_y = 0 \,.
\eea
\end{subequations}
It is easy to verify that these equations imply the Maxwell
equations (\ref{eq:maxwell-eqs-2}), thus providing a consistency check.
The fluctuation equations may be separated out to obtain higher
order equations for the different modes. One can verify that $A_x$
and $A_y$ satisfy the same fifth order equation.

We need to solve the equations (\ref{eq:einstein}) with
the condition that the solution satisfies ingoing wave
boundary conditions at the horizon $z{=}1$.
Near the horizon, the solutions behave as $A_{x,y}\sim (1{-}z)^\nu$,
$G_{x,y}\sim (1{-}z)^\beta$.
The set of equations (\ref{eq:einstein}) then determines the ingoing
and outgoing exponents
$\nu=\pm iw/(h^2{+}q^2{-}3)$, $\beta=1{+}\nu$.
The ingoing wave solution at the horizon
corresponds to $\nu_+=iw/(h^2{+}q^2{-}3)$.
Thus we will be looking for the solution of the form
\begin{subequations}
\label{eq:out-ansatz}
\begin{eqnarray}
  && A_x(z) = f(z)^{\nu_+}\; a_x(z)\,, \\
  && G_x(z) = f(z)^{1+\nu_+}\; g_x(z)\,,
\end{eqnarray}
\end{subequations}
and similarly for $A_y(z)$, $G_y(z)$. The functions $a_x(z)$,
$g_x(z)$ are required to be regular at the horizon $z{=}1$. There
is also a third possible exponent at the horizon, which leads to a
constant $z$-independent solution of equations (\ref{eq:einstein}),
\begin{equation}
\label{eq:constant-solution}
   G_y = \frac{iw}{h} A_x\,, \ \ \
   G_x =-\frac{iw}{h} A_y\,.
\end{equation}
This constant solution will be important later.

\subsection{Hydrodynamic limit}
\label{sec:hydro-limit}
\noindent
We are interested in the low frequency, hydrodynamic, behaviour of
the system when $\omega/T\ll\mu/T,B/T^2$. This regime of small
frequencies can be achieved by letting $\wn{\to}0$, with $q$ and
$h$ fixed. Therefore we will solve the equations perturbatively in
$\wn$
\begin{eqnarray}
  && a_x(z) = a_x^{(0)}(z) + w\, a_x^{(1)}(z) +\dots\,,\\
  && g_x(z) = g_x^{(0)}(z) + w\, g_x^{(1)}(z) +\dots\,,
\end{eqnarray}
and similarly for $a_y(z)$, $g_y(z)$.
To zeroth order in $\wn$, we have%
\footnote{
   Here and below we give expressions for the $x$ components;
   the $y$ components are the same but with
   $x\leftrightarrow y$ and $h \leftrightarrow -h$.
}
\begin{eqnarray}
  && g_x^{(0)\prime\prime}(z) +
     2\frac{\psi'(z)}{\psi(z)}\, g_x^{(0)\prime}(z) = 0\,, \\
  && a_x^{(0)\prime}(z) + q\, g_x^{(0)}(z) = 0\,,
\end{eqnarray}
where $\psi(z)\equiv f(z)/z$.
A general solution is of the form
$g_x^{(0)\prime}(z)={\rm const}/\psi(z)^2$
and the condition of regularity on the horizon implies
\begin{equation}
  g_x^{(0)}(z) = \gamma_x\,,\ \ \
  a_x^{(0)}(z) = \alpha_x - \gamma_x qz\,,
\label{eq:soln0}
\end{equation}
where $\gamma_x$, $\alpha_x$ are integration constants.
To first order in $\wn$ we find
\begin{eqnarray}
  && g_x^{(1)\prime\prime}(z) +
     2\frac{\psi'(z)}{\psi(z)}\, g_x^{(1)\prime}(z)
     = {\cal G}_x^{(0)}(z)\,, \\
  && a_x^{(1)\prime}(z) + q\, g_x^{(1)}(z) = {\cal A}_x^{(0)}(z)\,,
\end{eqnarray}
where the functions ${\cal G}_x^{(0)}(z)$, ${\cal A}_x^{(0)}(z)$
on the right hand side depend on the zeroth order solution
(\ref{eq:soln0}). One finds that ${\cal G}_x^{(0)}(z) \psi(z)^2$
and ${\cal A}_x^{(0)}(z) f(z)$ are fourth order polynomials in
$z$. A general solution is of the form
\begin{equation}
  g_x^{(1)\prime}(z) =
    \frac{\rm const}{\psi(z)^2} +
    \frac{1}{\psi(z)^2} \int_0^z\!\! {\cal G}_x^{(0)}(u)\,\psi(u)^2 \,du\,,
\end{equation}
and demanding regularity of $g_x^{(1)}(z)$ on the horizon
now implies a relation between the integration constants:
\begin{equation}
  \alpha_y = \frac{\gamma_x h(h^2{+}q^2{-}3) + 3 \gamma_y q(1{+}h^2{+}q^2)}
             {4(h^2{+}q^2)} \,.
\end{equation}
With this relation, one finds that the solution which is regular at
the horizon is
\begin{subequations}
\label{eq:solution-first-order-in-w}
\begin{eqnarray}
 && g_x^{(1)}(z) = \widetilde\gamma_x
                  -i\!\! \int_0^z \frac{du}{\psi(u)^2}
                   \;\gamma_x P_5(u)\,,\\
 && a_x^{(1)}(z) = \widetilde\alpha_x - q\!\! \int_0^z g_x^{(1)}(u) du
                  -i\int_0^z \frac{du}{f(u)}
                    (\gamma_x Q_4(u) + \gamma_y Q_3(u)) \,,
\end{eqnarray}
\end{subequations}
where $P_5, Q_3, Q_4$ are polynomials in $u$, whose coefficients
(given in the appendix) depend on $q$ and $h$ only. The
integration constants $\widetilde\gamma_x$, $\widetilde\alpha_x$
can be absorbed into $\gamma_x$ and $\alpha_x$ respectively; as a
result the solution is characterised by the two constants
$\gamma_x$, $\gamma_y$. This is too few. We want a solution that
is characterised by four constants: the four boundary values of
the fields $G_x$, $G_y$, $A_x$, $A_y$. We must therefore add to
the solution (\ref{eq:out-ansatz}) the constant solution
(\ref{eq:constant-solution})\footnote{The constant term is often
not written explicitly in treatments of black hole hydrodynamics,
which only give expressions for the derivatives of fields, but it
is usually there when more than one field is involved. Without it,
one would not have sufficiently many free constants at the
boundary. Gauge invariant fluctuations will be purely ingoing at
the horizon.}
\begin{equation}
  A_x = \d_{x},\ A_y = \d_{y},\
  G_x = -\frac{i\wn}{h}\d_{y},\
  G_y =  \frac{i\wn}{h}\d_{x} \,,
\end{equation}
where $\d_{x,y}$ are constants.
Then for the boundary values $G_x^0\equiv G_x(z{=}0)$ etc., we find
\begin{eqnarray}
 && G_x^0 =-\frac{i\wn}{h}\d_{y}+\gamma_x \,,\ \ \ \ \ \
    G_y^0 = \frac{i\wn}{h}\d_{x}+\gamma_y \,,\\
 && A_x^0 = \d_{x} + \alpha_x(\gamma_x,\gamma_y)\,,\ \ \
    A_y^0 = \d_{y} + \alpha_y(\gamma_x,\gamma_y)\,.
\end{eqnarray}
We can now express integration constants in terms of the
boundary values of the fields%
\footnote{
   The expressions (\ref{eq:alphaconst}) have $O(\wn)$
   corrections. However, these subleading terms do not contribute
   to the correlators to first order in $\wn$.
}
\begin{subequations}
\label{eq:alphaconst}
\begin{eqnarray}
  \d_{x} = A_x^0 +
        \frac{G_y^0 h(h^2{+}q^2{-}3) - 3 G_x^0 q(1{+}h^2{+}q^2)}
             {4(h^2{+}q^2)} \,,\\
  \d_{y} = A_y^0 -
        \frac{G_x^0 h(h^2{+}q^2{-}3) + 3 G_y^0 q(1{+}h^2{+}q^2)}
             {4(h^2{+}q^2)} \,.
\end{eqnarray}
\end{subequations}
To sum up: the solution to first order in $\wn$ is
\begin{subequations}
\label{eq:Gx-Ax-solution}
\begin{equation}
  G_x(z) = -\frac{i\wn}{h}\d_{y} + f(z)^{1+\nu_+}
           \left[G_x^0+\frac{i\wn}{h}\d_{y}
           -i\wn G_x^0 \int_0^z\frac{du}{\psi^2(u)}P_5(u)\right]\,,
\label{eq:Gx-solution}
\end{equation}
\begin{eqnarray}
  A_x(z) = \d_{x} + f(z)^{\nu_+} \left[
           A_x^0 - \d_{x} - (G_x^0{+}\frac{i\wn}{h}\d_{y})qz +
           i\wn q\, G_x^0 \int_0^z\!\!
           \frac{du (z-u)}{\psi^2(u)} P_5(u)\right.
           \nonumber\\
           \left. -i\wn \int_0^z \frac{du}{f(u)}
           (G_x^0 Q_4(u) + G_y^0 Q_3(u)) \right] \, .
\label{eq:Ax-solution}
\end{eqnarray}
\end{subequations}

\subsection{The action}

The current correlators will be computed by differentiating the
action evaluated on the solution with respect to the boundary
values of the fields. We therefore need the renormalised action
(\ref{eq:I-ren}) to quadratic order in perturbations.
The quadratic action evaluated on shell is given by the boundary
term
\be\label{eq:action}
I_{\text{ren.}} = \lim_{z \to 0} \frac{2 L^2 \a}{\k^2_{4}}
\int\!\!dt d^2x \left[
\frac{f^{1/2}-1}{2 z^3 f^{1/2}}
{\bf G} {\cdot} {\bf G} + \frac{q}{2} {\bf A} {\cdot} {\bf G} -
\frac{1}{8 z^2} {\bf G} {\cdot} {\bf G}' + \frac{f}{2} {\bf A} {\cdot}
{\bf A'}
\right] \,.
\ee
In this expression ${\bf G}\equiv(G_x, G_y)$, ${\bf A}\equiv(A_x,A_y)$.
Any possible contribution to the action from a boundary term
at the horizon is neglected \cite{Son:2002sd}. The action is now
computed by expanding the solution (\ref{eq:Gx-Ax-solution}) near the
boundary $z{=}0$ and substituting into Eq.~(\ref{eq:action}).

The result is most cleanly expressed in terms of the Fourier
transformed modes
\be
A_x^0(t) =  \int_{-\infty}^{\infty} \frac{d\w}{2\pi}\,
A_x^0(\w) e^{- i \w t} \,,
\ee
and similarly for the other modes. The action can be written as
the sum of three terms
\be\label{eq:evalaction}
I_{\text{ren.}} =  \frac{\sqrt{2} N^{3/2}}{6 \pi} \left[I_{AA} + I_{AG} + I_{GG} \right] \,,
\ee
where to lowest order in $\w/T \to 0$ we have
\be\label{eq:AAaction}
I_{AA} = \frac{ i q}{2 h} \int\frac{d\w}{2\pi} d^2x \; \w \ep_{ab} A_a^0(\w)
A_b^0(-\w) \,,
\ee
where $\ep_{a b}$ is antisymmetric, with $\ep_{xy} = 1$. The $\ep$ tensor appears because
of the mixing of $x$ and $y$ coefficients in (\ref{eq:Gx-Ax-solution}).
The term coupling the metric and gauge potential is
\be
I_{AG} = \frac{-i 3(1{+}q^2{+}h^2)}{4 h} \int\!\! d^2x
\frac{d\w}{2\pi} \;
 \w \ep_{ab}
      A_a^0(\w) G_b^0(-\w)\,,
\ee
and finally, dropping a contact term,
\be
I_{GG} = \int\!\!d^2x\frac{d\w}{2\pi}
           \left[\frac{(-3+h^2+q^2)^2}{32 (h^2{+}q^2)} \d_{ab} +
           \frac{9 q (1 {+} h^2 {+} q^2)^2}{32 h (h^2{+}q^2)}
\ep_{ab} \right] i \w G_a^0(\w) G_b^0(-\w) \,.
\ee
When writing these expression for the action,
the boundary values for the fields $A_{x,y}^0$, $G_{x,y}^0$
in the solution (\ref{eq:Gx-Ax-solution})
are taken as arbitrary functions of $\w$.
Taking functional derivatives with respect to the
boundary values allows us to compute hydrodynamic
correlators and conductivity.

\section{Hall conductivity and hydrodynamic correlators}

Recall from our discussion in section 2 that the background
magnetic field in the field theory is $B = h \a^2$ and the
charge density of the system $\rho \propto q \a^2$ is given by
(\ref{eq:densities}). These definitions imply that $\rho$ and $B$
have mass dimension two, which is the correct dimensionality for
charge density and field strength in three dimensions.

The AdS/CFT dictionary \cite{Son:2002sd} allows us to read off the large $N$
retarded Greens function from (\ref{eq:AAaction}) as
\begin{equation}
\label{eq:jj}
G^R_{ab}(\w) = -i\!\int\!\!d^2\!x\, dt\, e^{i \w t}
\theta(t) \langle [J_a(t), J_b(0)] \rangle =
 -i\w\ep_{ab} \frac{\rho}{B} \,.
\end{equation}
The conductivity is then given by the Kubo formula
(\ref{eq:Kubo-formula})
\begin{equation}
\boxed{
\begin{array}{ccccc}
\sigma_{xy} & = & - \sigma_{yx} & = &  {\displaystyle \frac{\rho}{B}} \,, \\
\sigma_{xx} & = & \sigma_{yy} & = & 0 \,.
\end{array}
}
\label{eq:sigma}
\end{equation}
There is no temperature dependence in this result. This expression
exactly recovers the results expected on general grounds from
Lorentz invariance (\ref{eq:classical}).

We can also compute from (\ref{eq:evalaction}) the retarded
correlator between the momentum density and the R charge current.
We have, to leading order in $\wn$ and at zero spatial
momentum,
\begin{eqnarray}
\label{eq:jt}
 G_{a\,\pi_b}^R(\w) =
 -i\! \int\!\!d^2\!x\, dt\, e^{i \w t} \theta(t)
      \langle [J_a(t), T_{tb}(0)] \rangle 
& = & - \frac{3 \e}{2 B} i \w \ep_{ab}
\,.
\end{eqnarray}
When extracting the correlators (\ref{eq:jt}),
the action $I_{AG}$ has to be multiplied by $\alpha$
because $G_x = \alpha g_{tx} (z/\alpha)^2$.
The final result here has been expressed in terms of
the background magnetic field and the charge and energy densities
of the equilibrium field theory. The expression for the
momentum-momentum correlator also follows from
(\ref{eq:evalaction}) as
\begin{eqnarray}
\label{eq:tt}
  G_{\pi_a\, \pi_b}^R(\w) & = &
  -i\! \int\!\!d^2\!x\, dt\, e^{i \w t} \theta(t)
  \langle [T_{ta}(t), T_{tb}(0)] \rangle \nonumber \\
& = & \frac{\sqrt{2} N^{3/2}}{6 \pi}  \frac{s^2 T^2}{\rho^2+\tilde B^2} i \w \d_{ab} - \frac{9 \, \rho \, \e^2}{4 B
(\rho^2+\tilde B^2)} i \w \ep_{ab} \,.
\end{eqnarray}
As previously, $\tilde B = B \sqrt{2} N^{3/2}/6\pi$. The
expressions (\ref{eq:jj}), (\ref{eq:jt}), and (\ref{eq:tt}) are
the main result of this paper. The regime of their applicability
is broader than the naive hydrodynamic limit $\w{\ll}T$. We have
assumed that $\w{\ll}\alpha$, with $\alpha$ implicitly defined by
the relation $4\pi T=\alpha(3-B^2/\alpha^4-\mu^2/\alpha^2)$. This
allows $\w/T$ to take any value provided the chemical potential
(or magnetic field) is sufficiently large.

\section{Discussion}

We have shown how a background magnetic field may be incorporated
into the AdS/CFT correspondence for 2+1 dimensional boundary
theories, by considering a dyonic black hole as the bulk
spacetime. As an application, we studied low frequency
charge transport in 2+1 dimensional CFTs 
whose gravity duals contain Einstein-Maxwell theory on $AdS_4$.
This class of conformal field theories includes
maximally supersymmetric $SU(N)$ Yang-Mills theory
at the conformal fixed point in the limit of large $N$.
For the Hall conductivity, our bulk computation recovered
the field theory result expected
due to Lorentz covariance.
In addition to the conductivity, we computed hydrodynamic response
functions of charge and momentum currents at zero spatial
momentum. One expects that the full study of bulk fluctuations
with nonzero spatial momentum should reproduce linearised
relativistic magnetohydrodynamics of the boundary field theory.
These computations appear to be technically involved, and we have
left them for future work.

From the fact that the diagonal d.c. conductivity
vanishes in (\ref{eq:sigma}) it follows that
the hydrodynamic limit $(\w/T){\to}0$
does not commute with the limit of small magnetic fields
$(B/T^2){\to}0$.
It would be interesting to understand the crossover
between the two regimes for the class of strongly
interacting CFTs studied here.

Let us also note that turning on a background magnetic field is
not the only way to introduce off diagonal conductivity in the
AdS/CFT framework. A topological $\theta$ term for the four
dimensional gauge fields in the bulk gives rise to the
antisymmetric contribution $\epsilon_{\mu\nu\lambda} p^\lambda$ in
two point current-current correlators of the 2+1 dimensional CFT
on the boundary
\cite{Witten:2003ya}.
Application of the Kubo formula now tells us that there is a
nonzero off diagonal conductivity, proportional to $\theta$.
However, introducing a $\theta$ term in the bulk does not
correspond to introducing a background magnetic field in the
original CFT; rather, it means that one simply studies a different
CFT \cite{Witten:2003ya}. Indeed, the effective Hall-like
conductivity coming from the $\theta$ term is a constant, present
even at zero temperature and zero charge density. In contrast,
Hall conductivity due to the background magnetic field studied in
this paper should be a nontrivial function of $\omega/T$ and
$\rho/T$. As a related application, it would be interesting to see
how a $\theta$ term may be obtained in $AdS_4$ via reduction from
ten or eleven dimensional supergravity. These terms seem to be
generic in flux compactifications.

There are various other interesting phenomena that arise in 2+1 dimensional
theories with a background magnetic field. These range from the
quantum Hall effect to the Nernst effect in superconductors. The
AdS/CFT correspondence provides a unique framework in which such
effects may be analytically studied in a strongly coupled field
theory.

\section*{Acknowledgements}

It is a pleasure to thank David Berenstein, Rob Myers, Subir
Sachdev, and Andrei Starinets for helpful conversations in the
course of this work. We would also like to thank Chris Herzog for
important comments on the preprint version of this text. This
research was supported in part by the National Science Foundation
under Grant No. PHY05-51164.

\appendix

\section{Polynomials}
The polynomials in (\ref{eq:solution-first-order-in-w}) are given by
\begin{eqnarray}
  P_5(u) &=&
     (7 h^2 {+} 7 q^2 {-} 3)                                  (1{-}u)^2
     -\frac{3+15(h^2{+}q^2)^2 - 22(h^2{+}q^2)}{h^2{+}q^2{-}3} (1{-}u)^3
     \nonumber\\
     &+&
      \frac{13(h^2{+}q^2)^2 - 7 (h^2{+}q^2)}{h^2{+}q^2{-}3}    (1{-}u)^4
     -\frac{4 (h^2{+}q^2)^2}{h^2{+}q^2{-}3}                    (1{-}u)^5\,,
\end{eqnarray}

\begin{equation}
  Q_3(u)=
  \frac{h(q^2{+}h^2{-}3)}{2(h^2{+}q^2)} (1{-}u)
  -
  \frac{3h(3h^2{+}3q^2{-}1)}{4(h^2{+}q^2)} (1{-}u)^2
  +
  h (1{-}u)^3\,,
\end{equation}

\begin{eqnarray}
  Q_4(u) &=&
  \frac{q(5q^2{+}5h^2{-}3)}{2(h^2{+}q^2)} (1{-}u)
   -
  \frac{3q(3 + 11(q^2{+}h^2)^2 - 18(q^2{+}h^2))}
       {4(q^2{+}h^2)(q^2{+}h^2{-}3)}
  (1{-}u)^2 \nonumber\\
  &+&
  \frac{2q(5q^2{+}5h^2{-}3)}{q^2{+}h^2{-}3} (1{-}u)^3
  -
  \frac{4q(q^2{+}h^2)}{q^2{+}h^2{-}3} (1{-}u)^4 \,.
\end{eqnarray}

\end{document}